\documentclass{article}
\usepackage{graphicx}
\usepackage{psfrag}
\begin{document}
\title{Cosmic Force}\author{Richard Drociuk}\date{\today}\maketitle
\section{Abstract}
The geodesics for classical particles in the gravitational field of a Schwarzschild black hole immersed in a vacuum of non-zero $\Lambda$ are found in their functional form. In particular, when r is a function of $\phi$, the functional form is in terms of quadrupoly periodic double theta functions. The turnig points are branch places on a one complex dimensional dissected Riemann surface. The four periods are found by integrating across the the period loops. It is shown that when $\Lambda$ is equal to zero, the quadrupoly periodic functions for the geodesics reduce to the functional form of doubly periodic elliptic functions.\\
\section{Introduction}

\indent{Steven Weinberg,[Weinberg,1],refered to the cosmological constant problem as crisis in physics, due to the discepency of some 120 orders of magnitude between the quantum field theory calculation and the cosmological calculation of the energy density in vacuum[Carroll,Ostlie,2]. Some progress is being made in astronomy with regards to this problem. The findings from the Type Ia Supernova data,[Filippenko,3], has created a new cosmology which is as revolutionary as Hubble's finding's of an expanding universe. Their observations suggest that the expansion of the universe is accellerating}\\
\indent{Presently there several investigations in progress[Efstathiou,4] which  attempt to measure cosmic accelleration, also known as Dark Energy. If this is to be modelled by a cosmic force, $F = \frac{1}{3}~\Lambda~r~m~c^2$ where c is the speed of light then the observations are in favor of a positive cosmological constant,$\Lambda$. This constant will be a fundamental constant of nature, and the cosmic force will be a fifth fundamental force of nature. Since the force is proportional to r, the method which probes the greatest distances will work the best, such as looking at anisotropies in the cosmic background radiation[Efstathiou,4]. The effects of the cosmological constant are neglible on the scale of the solar system[Neupane,5][Cardona,6]. It is vital to find independent experiments to mearsure cosmic accelleration[Boughn,Crittenden,7]. The equations in this paper may be tested on the scale of our galaxy, thus offering an independent measurement of the cosmological constant, in relatively well known setting. For example the work done  with galactic velocity rotation curves[Whitehouse,8] To do this consider the nucleus of the galaxy to be a black hole of radius $r_{g}=2GM/c^{2}$, where M is the estimated mass of the  galatic nucleus, and let the orbiting particle be a star or globular cluster. With such an experiment, one will have to make a reasonable estimate of the dark matter present. The mounting evidence in support of a cosmologicl constant makes it is necessary to know the mathematical relations and the geometry associated with an objects motion in space of intrinsic curvature and gravity. The developement of the mathematical formula that detemine the geodesics where formulated mostly by Riemann and Weierstrass, at least the general case[Baker,9,10].}\\  
\indent{To calculate the geodesics in the centrally symmetric gravitational field of a Scharzschild black hole in a universe with a cosmological constant, $\Lambda\neq 0$, the Hamilton-Jacobi method is applied to the metric,[Witten,11],[Hawking,12],[Rindler,13]}
\begin{equation}\label{1}
ds^{2}=g_{tt}~{dt}^{2}+g_{rr}~{dr}^{2}+g_{\phi\phi}~{d\phi}^{2}+g_{\theta\theta}~{d\theta}^{2}
\end{equation}
\begin{displaymath}
g_{tt} = (1-{\frac{1}{3}}{\sigma}~{\Lambda}~r^{2}-r_{g}/r)~c^{2}
\end{displaymath}
\begin{displaymath}
g_{rr} = -(1-{\frac{1}{3}}{\sigma}~{\Lambda}~r^{2}-r_{g}/r)^{-1}
\end{displaymath}
\begin{displaymath}
g_{\phi\phi} = -sin^{2}{\theta}~ r^{2}
\end{displaymath}
\begin{displaymath}
g_{\theta\theta} = -r^{2}
\end{displaymath}
Where $\sigma =\pm~1$ depending on whether or not the universe is de
Sitter or anti-de Sitter, respectively. $\Lambda$ is the cosmological
constant and was first introduced by Einstein[Filippenko,3] and $r_{g}$ was
introduced by Schwarzschild. The functional form of the geodesics when
$r_{g}=0$ have been studied, they are given in terms of genus zero
circular functions.  When $\Lambda=0$ the geodesics are given by genus one
elliptic functions [Bartlett,14]. In this paper, the functional form of
the geodesics when $\Lambda$ and $r_{g}$ are both non-zero are  found,
in terms of Riemann's hyperelliptic theta functions[Baker, 9,10].
\section{ Derivation of the Hyper-elliptic $\phi$ Integral}
\indent{Let a particle of mass $m$ be constrained to move in the equitorial plane, $\theta = 0$, and let the action be given by [Landau,15], 
\begin{equation}\label{2}
\ S=-E~t+L~\phi+S_r
\end{equation}
where $E$ is the total energy of the particle and $L$ is it's angular momentum. By the conservation of four momentum,
\begin{equation}\label{3}
\ g^{ij}~\frac{dS}{dx^{i}}~\frac{dS}{dx^{j}}= m^{2}
\end{equation}
\ we can insert $S$ and $g^{i,j}$ from (\ref{1}) and (\ref{2})into (\ref{3}) and obtain,
\begin{equation}\label{4}
\ S_{r} = \int_{r_0}^{r}~\frac{(E^{2}-(1-\frac{1}{3}~\Lambda~\sigma~r^{2}-r_{g}/r)~(L^{2}/r^{2}+m^{2}~c^{2}))^{\frac{1}{2}}}{(1-\frac{1}{3}~\Lambda~\sigma~r^{2}-r_{g}/r)}~dr
\end{equation}
Differentiating $S_{r}$ w.r.t. $L$ and using
\begin{equation}\label{5}
\ \phi + \frac{\partial~S_{r}}{\partial~L} = const = 0
\end{equation}
\begin{equation}\label{6}
\ \phi = \int_{r_{0}}^{r}\frac{L~dr}{r^{2}~(E^{2}-(1-\frac{1}{3}~\Lambda~\sigma~r^{2}-r_{g}/r)~(L^{2}/r^{2}+m^{2}~c^{2}))^{\frac{1}{2}}}
\end{equation}
This integral is a hyperelliptic integral of genus, $g=2$. The problem
is to invert this integral, i.e. express the upper limit r as a
function of $\phi$ ,[Jacobi,16],[Gauss,17],[Abel,18], this is known as the inversion problem.}

\section{The Solution to the Hyper-elliptic Integral of Arbitrary Genus}
\indent{The following will be a review of the account given by H.F. Baker[9,10] and the references contained in these books. The general solution to the inversion problem for hyperelliptic integrals is given by solving $g$ of the $2~g+1$ equations, for the $g$ variable places $x_{1},\ldots,x_{g}$,
\begin{equation}\label{7}
\ \frac{\vartheta^{2}(u|u^{b,a})}{\vartheta^{2}(u|u^{b',a'})}= A(b)~(b-x_{1})\ldots(b-x_{g})
\end{equation}
The notation Baker uses for the generalized theta function is,
\begin{equation}\label{8}
\ \vartheta(u|u^{b,a})= \vartheta(u|\frac{1}{2}~\Omega_{m,m'})=\vartheta(u;\frac{1}{2}~m,\frac{1}{2}~m')= e^{a~u^{2}}~\Theta(v;\frac{1}{2}~m,\frac{1}{2}~m')
\end{equation}
where $a$ is an arbitrary $g\times g$  symetrical matrix, since  putting  (\ref{8})  into (\ref{7})  gives,
\begin{equation}\label{9}
\ \frac{\vartheta^{2}(u|u^{b,a})}{\vartheta^{2}(u|u^{b',a'})}= \frac{\Theta^{2}(v;\frac{1}{2}~m,\frac{1}{2}~m')}{\Theta^{2}(v;\frac{1}{2}~k,\frac{1}{2}~k')}
\end{equation}
where $m, m', k$ and $k'$ are integers. Riemann's theta functions are defined as,
\begin{equation}\label{10}      
\Theta(v;\frac{1}{2}~m,\frac{1}{2}~m')=\Sigma~e^{2~h~v~(n+\frac{1}{2}~m')+b~(n+\frac{1}{2}~m')^{2}+i~\pi~m~(n+\frac{1}{2}~m')}
\end{equation}
with, $\Sigma = \sum_{n_1=-\infty}^{+\infty}\ldots\sum_{n_{g}=-\infty}^{+\infty}$. $h$ is a $g\times g$ matrix,in general non-symmetrical. $b$ is a $g \times g$ symmetrical matrix.}
In (\ref{9}), u denotes the $g$ quantities,
\begin{equation}\label{11}
u_{i}^{x_{1},a_{1}}+\ldots+u_{i}^{x_{g},a_{g}} = u_{i} 
\end{equation}
with, $i=1\ldots g$. This is known as Abel's Theorem. Riemann's Normal Integral of the first kind is defined as,
\begin{equation}\label{12}
u_{i}^{x,a}=\int_{a}^{x}\frac{(x,1)_{i,g-1}~dx}{y} 
\end{equation}
where
\begin{equation}\label{13}
(x,1)_{i,g-1}=A_{i,g-1}~x^{g-1}+A_{i,g-2}~x^{g-2}+\ldots+A_{i,0}
\end{equation}
and $A_{i,g-1},\ldots,A_{i,0}$ are arbitrary constants. The denominator in (\ref{12}) is given by,
\begin{equation}\label{14}
y^{2}=4~(x-a_{1})\ldots(x-a_{g})~(x-c_{1})\ldots(x-c_{g})~(x-c)
\end{equation}
where $a_{1},\ldots,a_{g},c_{1},\ldots,c$ are branch places and $x$ is a variable on the one complex dimensional Riemann surface of genus, $g$, shown in figure 1.
\\
\begin{figure}[tbp]
\begin{center}
\psfrag{x}{$\omega_{r,1}$}
\psfrag{xp}{$\omega'_{r,1}$}
\psfrag{y}{$\omega_{r,2}$}
\psfrag{yp}{$\omega'_{r,2}$}
\psfrag{g}{$\omega_{r,g}$}
\psfrag{gp}{$\omega'_{r,g}$}
\psfrag{z1}{$\alpha_{1}$}
\psfrag{z2}{$\alpha_{2}$}
\psfrag{zg}{$\alpha_{g}$}
\psfrag{z3}{$\beta_{1}$}
\psfrag{z4}{$\beta_{2}$}
\psfrag{zgg}{$\beta_{g}$}
\psfrag{a_1}{$a_{1}$}
\psfrag{c_1}{$c_{1}$}
\psfrag{a_2}{$a_{2}$}
\psfrag{c_2}{$c_{2}$}
\psfrag{a_g}{$a_{g}$}
\psfrag{c_g}{$c_{g}$}
\psfrag{c}{$c$}
\psfrag{a}{$a$}
\includegraphics{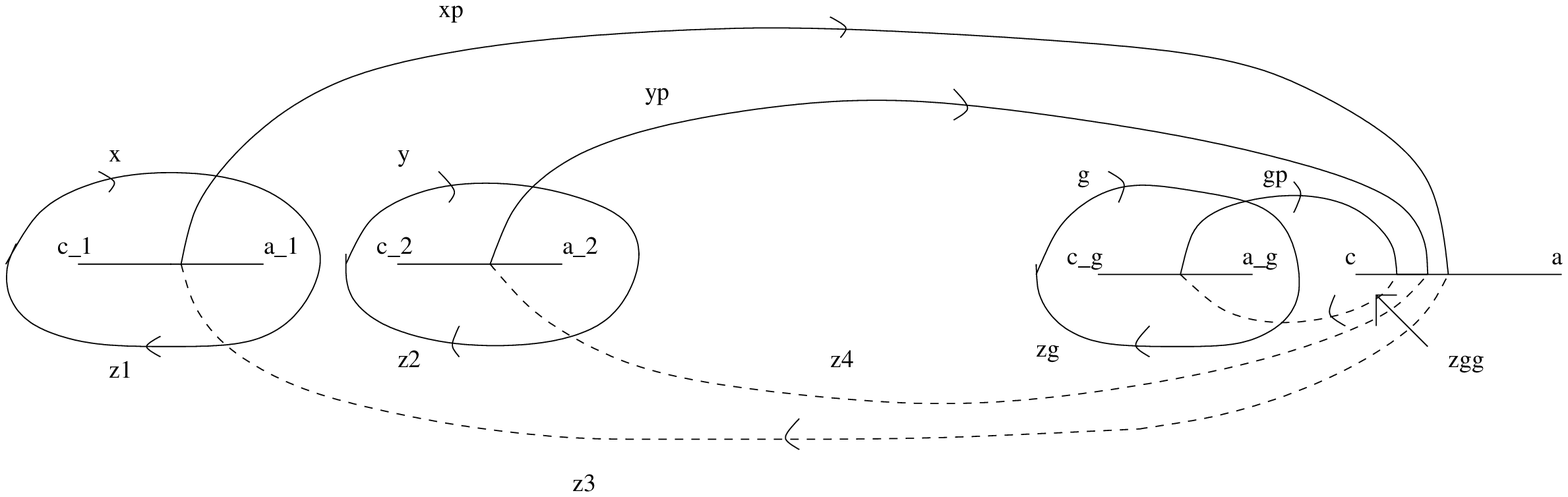}
\caption{\label{fig:1}}
One Complex Dimensional Dissected Arbitrary Genus Riemann Surface
\end{center}
\end{figure}
From figure(\ref{fig:1}) the period table, table 1, is derived.\\
\\
\\Table 1. Primative Normal Periods[Brioschi,19][Bolza,20][Konigsberger,21]\\
\\
\begin{tabular}{|r|l|l|r|l|r|l|}
\hline
Int & $\alpha_{1}$ & \ldots & $\alpha_{g}$ & $\beta_{1}$ &\ldots& $\beta_{g}$ \\
\hline
$u_{1}^{x,a}$&$\omega_{11}$&\ldots&$\omega_{1g}$&$\omega'_{11}$&\ldots&$\omega'_{1g}$\\
\hline
\ldots&\ldots&\ldots&\ldots&\ldots&\ldots&\ldots\\
\hline
$u_{g}^{x,a}$&$\omega_{g1}$&\ldots&$\omega_{gg}$&$\omega'_{g1}$&\ldots&$\omega'_{gg}$\\
\hline
\end{tabular} \\
\\
Now let,
\begin{equation}\label{15}
\frac{1}{2}~\Omega_{m,m'} = m_{1}~\omega_{r,1}+\ldots+m_{g}~\omega_{r,g}+m'_{r,1}~\omega'_{r,1}+\ldots+m'_{g}~\omega'_{r,g}
\end{equation}
where $m$ and $m'$ are equal to +1 when integrating across a period loop from right to left, -1 when crossing from left to right, and 0 when not crossing a period loop.
Now define an integral of the first kind via,
\begin{equation}\label{16}
\ \pi~i~v_{r}^{x,a}=h_{r,1}~u_{1}^{x,a}+h_{r,2}~u_{2}^{x,a}+\ldots +h_{r,g}~u_{g}^{x,a}
\end{equation}
with $r = 1,2,  \ldots,g$ and where,
\begin{equation}\label{17}
\ 2~h~w = \pi~i
\end{equation}
\begin{equation}\label{18}
\ b=2~h~w' = \pi~i~\tau
\end{equation}
so with these definitions table 1 becomes table 2, below
\\
\\Table 2. Hyperelliptic Moduli\\
\\
\begin{tabular}{|r|l|r|l|r|l|r|l|}
\hline
Int& $\alpha_1$ & $\alpha_2$ & \ldots& $\alpha_g$ & $\beta_1$ &\ldots& $\beta_g$ \\
\hline
$v_{1}^{x,a}$&$1$&$0$&\ldots&$0$&$\tau_{1,1}$&\ldots&$\tau_{1,g}$\\
\hline
\ldots&\ldots&\ldots&\ldots&\ldots&\ldots&\ldots&\ldots\\
\hline
$v_{g}^{x,a}$&$0$&$0$&\ldots&$1$&$\tau_{g,1}$&\ldots&$\tau_{g,g}$\\
\hline
\end{tabular}\\
\\
with Riemann's condition for hyperelliptic integrals,[Brioschi,13],[Bolza,14],
\begin{equation}\label{19}
 \tau = \tau^{t}
\end{equation}
With this condition there are $\frac{1}{2}~n~(n+1)+1$ different non-zero periods. With the above equations, Riemann's  theta function becomes,
\begin{equation}\label{20}
 \Theta(v;\frac{1}{2}~m,\frac{1}{2}~m')=\Sigma~e^{2~\pi~i~v~(n+\frac{1}{2}~m')+i~\pi~\tau~(n+\frac{1}{2}~m')^{2}+i~\pi~m~(n+\frac{1}{2}~m')}
\end{equation}
where the quadratic forms are given by,
\begin{equation}\label{21}
 v~(n+\frac{1}{2}~m')=v~n+\frac{1}{2}~m'~v=v_{1}~n_1+\ldots+v_{g}~n_{g}+\frac{1}{2}~v_{1}~m'_{1}+\ldots+\frac{1}{2}~v_{g}~m'_{g}
\end{equation}
and similarly, 
\begin{eqnarray}\label{22}
 \tau~(n+\frac{1}{2}~m')^2 & = &\tau~n^{2}+\tau~n~m'+\frac{1}{4}~\tau~m'^{2}\nonumber\\ & &
=(\tau_{1,1}~n_{1}^2+2~\tau_{1,2}~n_{1}~n_{2}+\ldots+\tau_{g,g}~(n_{g})^{2})\nonumber\\ & & +\sum_{s=1}^{g}\sum_{r=1}^{g}~(\tau_{r,s})~n_{r}~m'_{s}\nonumber\\ & & +\frac{1}{4}~(\tau_{1,1}~(m'_{1})^{2}+2~\tau_{1,2}~m'_1~m'_2+\ldots+\tau_{g,g}~(m'_{g})^{2})
\end{eqnarray}
It has been proven by Riemann that for the integral of the firt kind, i.e. $v_{r}^{x,a}$, when it has the period scheme as in table 2, then the imaginary part of,
\begin{equation}\label{23}
 \tau_{1,1}~n_{1}^2+2~\tau_{1,2}~n_{1}~n_{2}+\ldots+\tau_{g,g}~n_{g}^{2}
\end{equation}
is positive for all integer values of $n_{1}$ and $n_{2}$, with the exception of $n_{1}=n_{2}=0$. Therefore the modulus of $e^{i~\pi~\tau~n^{2}}$ is less than unity and the function(\ref{20}) converges for all values of it's argument, v.
The last thing to define in (\ref{7}) is the constant $A(b)$, it is defined as,
\begin{equation}\label{24}
 A(b) = (\epsilon~\frac{d}{dx}~(x-a_{1})\ldots(x-a_{g})~(x-c_1)\ldots(x-c_{g})(x-c))|_{x=b})^{-\frac{1}{2}}
\end{equation} 
where, $\epsilon$ is +1 when $u^{b,a}$ is an odd half period and -1 when it is an even half period.
\section{The Solution to the  Hyperelliptic Integral of Genus Two}
The solution of the hyper-elliptic integral(\ref{6}), is given by(\ref{7}),[Rosenhain,22],[Gopel,23],[Forsyth,24], [Baker,9], with $g=2$ we have
\begin{equation}\label{25}
\ \frac{\Theta^{2}(v_{1},v_{2};\frac{1}{2}~m_{1},\frac{1}{2}~m_{2},\frac{1}{2}~m'_{1},\frac{1}{2}~m'_{2})}{\Theta^{2}(v_{1},v_{2};\frac{1}{2}~k_{1},\frac{1}{2}~k_{2},\frac{1}{2}~k'_{1},\frac{1}{2}~k'_{2})}= A(b)~(b-x_{1})~(b-x_{2})
\end{equation}
where by(\ref{16})we have,[Baker,9,10],
\begin{eqnarray}\label{26}
 \pi~i~v_{1}^{x,a} & = & h_{1,1}~u_{1}^{x,a} + h_{1,2}~u_{2}^{x,a}\nonumber\\ 
 \pi~i~v_{2}^{x,a} & = & h_{2,1}~u_{1}^{x,a} + h_{2,2}~u_{2}^{x,a}\ 
\end{eqnarray}
where $h_{i,j}$ are the elements of the following matrix,
\begin{displaymath}
\mathbf{h} = \frac{\pi~i}{2~\Delta}~
\left( \begin{array}{ccc}
\omega_{2,2} & -\omega_{1,2}\\
-\omega_{2,1}&\omega_{1,1}\\
\end{array}\right)
\end{displaymath}
and,
 \begin{equation}\label{27}
\Delta = det(h)\neq 0
\end{equation}
Abel's Theorem(\ref{11}), becomes
\begin{eqnarray}\label{28}
\ u_{1}^{x_{1},a_{1}}+ u_{1}^{x_{2},a_{1}} & = & u_{1}\nonumber\\
\ u_{2}^{x_{1},a_{1}}+ u_{2}^{x_{2},a_{1}} & = & u_{2}\
\end{eqnarray} 
these become with,
 \begin{equation}\label{29}
\ x_{2} = a_{2}
\end{equation}
 and with (\ref{12}),
\begin{eqnarray}\label{30}
\ u_{1}^{x_{1},a_{1}}  & = & u_{1} \nonumber\\ 
                       & = &\int_{a_{1}}^{x_{1}}\frac{(A_{1,0}+A_{1,1}~x)~dx}{y}\nonumber\\
\ u_{2}^{x_{1},a_{1}}  & = & u_{2} \nonumber\\
                       & = &\int_{a_{1}}^{x_{1}}\frac{(A_{2,0}+A_{2,1}~x)~dx}{y} \nonumber\\
\end{eqnarray}
where, 
 \begin{equation}\label{31}
\ y^2 = 4~x^5 + \lambda_{4}~x^{4} +\lambda_{3}~x^{3} +\lambda_{2}~x^{2} +\lambda_{1}~x + \lambda_{0}
\end{equation}
These coeffients may be substituted into the general solution to the quintic equation [Drociuk,25] where the branch places, i.e. the roots of the quintic(\ref{31}), are determined in their functional form. Then (\ref{31})maybe rewritten in the form,
\begin{equation}\label{32}
\ y^2 = 4~(x-a_{1})~(x-a_{2})~(x-c_{1})~(x-c_{2})~(x-c)
\end{equation}
So the the Riemann Surface, figure 1, becomes a genus 2 Riemann surface figure 2.
\\
\\
\begin{figure}[tbp]
\begin{center}
\psfrag{x}{$\omega_{r,1}$}
\psfrag{xp}{$\omega'_{r,1}$}
\psfrag{y}{$\omega_{r,2}$}
\psfrag{yp}{$\omega'_{r,2}$}
\psfrag{g}{$\omega_{r,g}$}
\psfrag{gp}{$\omega'_{r,g}$}
\psfrag{z1}{$\alpha_{1}$}
\psfrag{z2}{$\alpha_{2}$}
\psfrag{zg}{$\alpha_{g}$}
\psfrag{z3}{$\beta_{1}$}
\psfrag{z4}{$\beta_{2}$}
\psfrag{zgg}{$\beta_{g}$}
\psfrag{a_1}{$a_{1}$}
\psfrag{c_1}{$c_{1}$}
\psfrag{a_2}{$a_{2}$}
\psfrag{c_2}{$c_{2}$}
\psfrag{a_g}{$a_{g}$}
\psfrag{c_g}{$c_{g}$}
\psfrag{c}{$c$}
\psfrag{a}{$a$}
\includegraphics{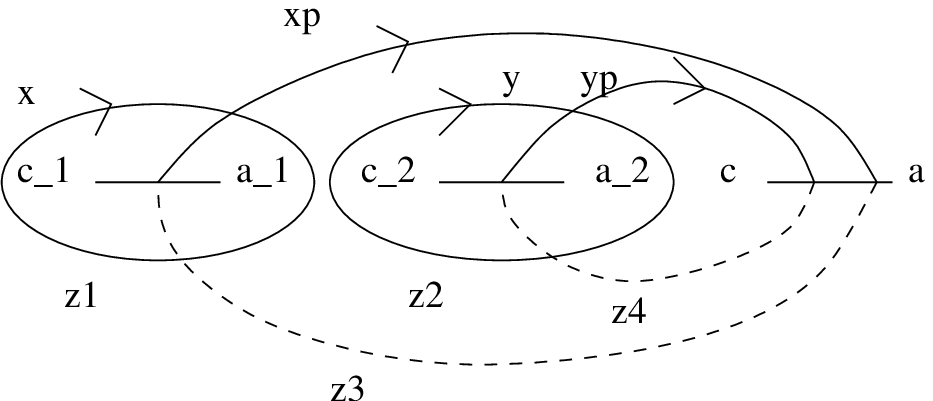}
\caption{\label{fig:2}}
Genus Two Riemann Surface
\end{center}
\end{figure}
From figure(\ref{fig:2}) the period table, table 3, is derived.\\
\\
\\Table 3. Period Table for Genus Two Hyperelliptic Integrals\\
\\
\begin{tabular}{|r|l|r|l|r|l|}
\hline
Int & $\alpha_{1}$ & $\alpha_{2}$ & $\beta_{1}$ & $\beta_{2}$ \\
\hline
$u_{1}^{x,a}$&$\omega_{11}$&$\omega_{12}$&$\omega'_{11}$&$\omega'_{12}$\\
\hline
$u_{2}^{x,a}$&$\omega_{21}$&$\omega_{22}$&$\omega'_{21}$&$\omega'_{22}$\\
\hline
\end{tabular} \\
\\
\\
With definitions(\ref{26}), table three becomes table four,
\\
\\Table 4. Genus Two Hyperelliptic Moduli\\
\\
\begin{tabular}{|r|l|r|l|r|l|}
\hline
Int& $\alpha_1$ & $\alpha_2$ & $\beta_1$ & $\beta_2$ \\
\hline
$v_{1}^{x,a}$&$1$&$0$&$\tau_{1,1}$&$\tau_{1,2}$\\
\hline
$v_{2}^{x,a}$&$0$&$1$&$\tau_{2,1}$&$\tau_{2,2}$\\
\hline
\end{tabular}\\
\\ 
So the four non-zero periods are, 1 and 
\begin{eqnarray}\label{33}
\ \tau_{1,1} & = & \frac{\omega_{2,2}~\omega'_{1,1}-\omega_{1,2}~\omega'_{2,1}}{\Delta} \nonumber\\
\ \tau_{2,2} & = & \frac{-\omega_{2,1}~\omega'_{2,2}+\omega_{1,1}~\omega'_{2,2}}{\Delta} \
\end{eqnarray}
and by Riemann's condition(\ref{19}),
  \begin{eqnarray}\label{34}
\ \tau_{1,2} & = & \frac{-\omega_{2,1}~\omega'_{1,1}+\omega_{1,1}~\omega'_{2,1}}{\Delta}\nonumber\\
& = & \frac{\omega_{2,2}~\omega'_{1,2}-\omega_{1,2}~\omega'_{2,2}}{\Delta}\nonumber\\
& = & \tau_{2,1}\
\end{eqnarray}
The quadrupoly periodic Double Theta Function is given is from(\ref{20}),(\ref{21}),(\ref{22}),
\begin{displaymath}\
\Theta(v_{1},v_{2};\frac{1}{2}~m_{1},\frac{1}{2}~m_{2},\frac{1}{2}~m'_{1},\frac{1}{2}~m'_{2})
\end{displaymath}
\begin{equation}\label{35}
 = \sum_{n_{1}=-\infty}^{\infty}\sum_{n_{2}=-\infty}^{\infty}~e^{\pi~i~(v_{1}~(2~n_{1}+m'_{1})+v_{2}~(2~n_{2}+m'_{2}))}\times
\end{equation}
\begin{equation}
 q^{(2~n_{2}+m'_{2})^{2}}~q'^{(2~n_{1}+m'_{1})^{2}}~r^{(2~n_{1}+m'_{1})~(2~n_{2}+m'_{2})}\times (-1)^{m_{1}~n_{1}+m_{2}~n_{2}}\times (i)^{m_{1}~m'_{1}+m_{2}~m'_{2}}
\end{equation}
where,
\begin{eqnarray}\label{36}
q & = & e^{\frac{1}{4}~i~\pi~\tau_{1,1}} \nonumber\\
q' & = & e^{\frac{1}{4}~i~\pi~\tau_{2,2}} \nonumber\\
r & = & e^{\frac{1}{2}~i~\pi~\tau_{1,2}} \
\end{eqnarray}
which is equal to Forsyth's double theta function,[Forsyth,24], with the two characteristics related by the following,
\begin{eqnarray}\label{37}
m_{1} & = & \rho   \nonumber\\
m_{2} & = & \rho'   \nonumber\\
m'_{1} & = & \sigma  \nonumber\\
m'_{2} & = & \sigma'  \
\end{eqnarray}
so that $a_{r}$ in Forsyth's paper is equal to
\begin{equation}\label{38}
a_{r} = (i)^{\sigma'~\rho'+\rho~\sigma}~r^{(2~n+\sigma')~(2~m+sigma)}
\end{equation}
where $n=n_{1}$ and $m=n_{2}$ are the summation indices. Forsyth gives a review of Rosenhain's theory, [Forsyth,26] of the fifteen ratio's of the quadrupoly periodic theta function. From the list given in [Forsyth,24], the first ratio is selected and is equal to,
\begin{equation}\label{39}
\frac{\Theta^{2}(v_{1},v_{2};\frac{1}{2}~(1),\frac{1}{2}~(1),\frac{1}{2}~(1),\frac{1}{2}~(0))}{\Theta^{2}(v_{1},v_{2};\frac{1}{2}~(1),\frac{1}{2}~(0),\frac{1}{2}~(0),\frac{1}{2}~(1))}=\sqrt{\frac{(a_{1}-a_{2})}{(a_{1}-c_{1})~(a_{1}-c_{2})~(a_{1}-c)}}~(x_{1}-c)
\end{equation}
where (\ref{24})(\ref{25})(\ref{29})and(\ref{35})have been used. The double theta funcntions used in (\ref{39}) are
\begin{eqnarray}\label{40}
\theta_{13} = \sum_{n_{1}=-\infty}^{\infty}~\sum_{n_{2}=-\infty}^{\infty}~e^{\pi~i~(v_{1}~(2~n_{1}+1)+v_{2}~(2~n_{2}))}~q^{(2~n_{2})^{2}}~{}\nonumber\\
~q'^{(2~n_{1}+1)^{2}}~r^{(2~n_{1}+1)~(2~n_{2})}~ (-1)^{n_{1}+n_{2}} ~i 
\end{eqnarray}
where Forsyth's $\theta$ notation is adopted,
\begin{equation}\label{41}
\theta_{13} = \Theta(v_{1},v_{2};\frac{1}{2},\frac{1}{2},\frac{1}{2},0)
\end{equation}
and in the denominator of (\ref{39}), is
\begin{eqnarray}\label{42}
 \theta_{12}= \sum_{n_{1}=-\infty}^{\infty}~\sum_{n_{2}=-\infty}^{\infty}~e^{\pi~i~(v_{1}~(2~n_{1})+v_{2}~(2~n_{2}+1))}~q^{(2~n_{2}+1)^{2}}~{}\nonumber\\
{}~q'^{(2~n_{1})^{2}}~r^{(2~n_{1})~(2~n_{2}+1)}~ (-1)^{n_{1}} 
\end{eqnarray}
\begin{equation}\label{43}
\theta_{12}= \Theta(v_{1},v_{2};\frac{1}{2},0,0,\frac{1}{2})
\end{equation}
\section{The Solution to the Hyper-elliptic $\phi$ Integral}
 The solution to the hyper-elliptic $\phi$ integral(\ref{6})is, with $ x_{1}=\frac{r_{g}}{r}$,
\begin{equation}\label{44}
 r = ( r_{g}~\theta^{2}_{12}~\sqrt{a_{1}-a_{2}})/(c~\sqrt{a_{1}-a_{2}}~\theta^{2}_{12}+\sqrt{(a_{1}-c_{1})~(a_{1}-c_{2})~(a_{1}-c)}~\theta^{2}_{13})
\end{equation}
where $\theta_{13}$ and $\theta_{12}$ are given in (\ref{40})and(\ref{42}). The hyper-elliptic $\phi$ integral(\ref{6})may be rewritten as,
\begin{equation}\label{45}
\int_{a_{1}}^{x_{1}}~\frac{x~dx}{y}=-\frac{\phi}{2}
\end{equation}
and define a new variable, $\phi'$, given by
\begin{equation}\label{46}
\int_{a_{1}}^{x_{1}}~\frac{dx}{y}=\frac{\phi'}{2}
\end{equation}
so that the variables in the argument of the theta functions are,
\begin{eqnarray}\label{47}
 v_{1} & = & ((h_{1,1}~A_{1,0}+h_{1,2}~A_{2,0})~\phi'-(h_{1,1}~A_{1,1}+h_{1,2}~A_{2,1})~\phi)/(2~\pi~i)   \nonumber\\ 
 v_{2} & = & ((h_{2,1}~A_{1,0}+h_{2,2}~A_{2,0})~\phi'-(h_{2,1}~A_{1,1}+h_{2,2}~A_{2,1})~\phi)/(2~\pi~i)  \
\end{eqnarray}
where {h} is defined in (\ref{26})and(\ref{27}). The constants are given by,
\begin{equation}
 A_{1,0}=-a_{1}=A_{2,0}
\end{equation}
\begin{equation}
 A_{1,1}=A_{2,1}=1
\end{equation}
With this choice the equations of Baker and Forsyth are identical to these, in the case of $g=2$.
 The roots of the quintic,$ a_{1}, a_{2}, c_{1}, c_{2}$ and $c$ are given by setting the coeffients in (\ref{31}) equal to,
\begin{equation}\label{48}
\lambda_{0}  = (4~\Lambda~m^{2}~c^{2}~r_{g}^{4})/(3~L^{2})
\end{equation}
\begin{equation}\label{49}
\lambda_{1} =  0
\end{equation}
\begin{equation}\label{50}
\lambda_{2} =  4~(E^{2}~c^{2}~L^{2}+\Lambda/3-m^{2}~c^{2})~r_{g}^{2}
\end{equation}
\begin{equation}\label{51}
\lambda_{3}  =  (4~m^{2}~c^{2}~r_{g}^{2})/L^{2}
\end{equation}
\begin{equation}\label{52}
\lambda_{4} = -4
\end{equation}
then substuting into the solution to the quintic [Drociuk, 25]. If one thinks in terms of the advancement of perhelia, i.e. Mercury, in the case $\Lambda=0$, the orbits obey doubly periodic functions. In the case when $\Lambda \neq 0$ they are equal to quadrupoly periodic functions of two variables $\phi$ and $\phi'$, equations (\ref{45}) and (\ref{46}). $\phi$ is the physical angular coordinate defined in equation (\ref{1}). $\phi'$ exist's because of Abel's theorem, equation (\ref{28}).Simply taking $\phi'$ equal to zero cannot work for then equation (\ref{27})will go to zero and the transformation (\ref{26}) becomes singular. Therefore $\phi'$ is an unknown parameter?  

\section{The Reduction of the Accellerating Universe to the General Relativistic Universe}
When the Cosmic Force is set to zero and particles are subject to gravitation alone, two roots of the quintic, see [Drociuk,25] or by inspection, go to zero. Then $\tau_{1,2}=\tau_{2,2}\Rightarrow 0$, then $q'=r=1$, and setting $n=0$, gives
\begin{equation}\label{53}
\frac{\theta^{2}_{13}}{\theta^{2}_{12}}\Rightarrow -k~sn^{2}(v_{1})
\end{equation}
Where sn is one of Jacobi's four elliptic functions [Forsyth, 26]. Choosing to let $a_{2}=c_{2}\Rightarrow 0$, then (\ref{44})becomes 
\begin{equation}\label{54}
-k~sn^2(v_{1}) = (a_{1}-x_{1})/\sqrt{(a_{1}-c_{1})~(a_{1}-c)}
\end{equation}
where,
\begin{eqnarray}\label{55}
a_1 & = & r_{g}/r_{1}  \nonumber\\
c_1 & = & r_{g}/r_{3}  \nonumber\\
c   & = & r_{g}/r_{5}  \nonumber\\
k   & = & \sqrt{(c_{1}-a_{1})/(c-a_{1})}  \
\end{eqnarray}
which gives the Schwarzschild black hole solution without the cosmological constant, as a genus 1 elliptic function, which is the same as,[Bartlett,14], it is
\begin{equation}\label{56}
(\frac{1}{r}-\frac{1}{r_{1}})=(\frac{1}{r_{3}}-\frac{1}{r_{1}})~sn^{2}(v_{1})
\end{equation}
Since two roots are zero, (\ref{32})becomes with $a_{2}=c_{2}=0$,
\begin{equation}\label{57}
y^2 = 4~(x-a_{1})~(x-c_{1})~(x-c)
\end{equation}
which gives rise to genus $g=1$ functions whose Riemann Surface is,
Figure 3
\\
\begin{figure}[tbp]
\begin{center}
\psfrag{x11}{$\omega_{1,1}$}
\psfrag{x11p}{$\omega'_{1,1}$}
\psfrag{z1}{$\alpha_{1}$}
\psfrag{z3}{$\beta_{2}$}
\psfrag{a_1}{$a_{1}$}
\psfrag{c_1}{$c_{1}$}
\psfrag{c}{$c$}
\psfrag{a}{$a$}
\includegraphics{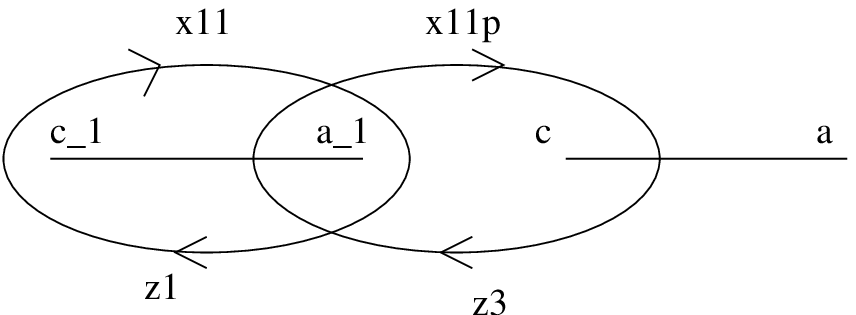}
\caption{\label{fig:3}}
Genus One Riemann Surface
\end{center}
\end{figure}
and the period table reduces to,\\
\\
\\
\\
\\
Table 5 Normal Primative Periods Genus One\\
\\
\begin{tabular}{|r|l|r|l|r|l|}
\hline
Int & $\alpha_{1}$ & $\beta_{1}$ \\
\hline
$u_{1}^{x,a}$&$\omega_{11}$&$\omega'_{11}$\\
\hline
\end{tabular}
\\
\\
\\
then by (\ref{26}), with $h_{1,2}=h_{2,1}=h_{2,2}=0$, gives
\begin{equation}\label{58}
h=h_{11}=\pi~i/\omega_{11}
\end{equation}
Then table 4 reduces to,\\
\\
Table 6  Genus One Moduli\\
\\
\begin{tabular}{|r|l|r|l|r|l|}
\hline
Int & $\alpha_{1}$ & $\beta_{1}$ \\
\hline
$v_{1}^{x,a}$&$1$&$\tau_{11}$\\
\hline
\end{tabular}
\\
\\
\\
Where,
\begin{equation}\label{59}
\tau_{11} = (\omega'_{11})/(\omega_{11})= (\int_{a_{1}}^{c_{1}}~dx/y)/(\int_{a_{1}}^{c}~dx/y)
\end{equation}
and
\begin{equation}\label{60}
\pi~i~v_{1}^{x,a}= -(\frac{\pi~i}{\omega_{11}}~\int_{a}^{x}~A_{1,1}~dx/y)
\end{equation}
which becomes with (\ref{45}),
\begin{equation}\label{61}
v_{1} = (A_{1,1}~\phi)/(\omega_{11}~2)
\end{equation}
and choose,
\begin{equation}\label{62}
A_{1,1} = \omega_{1,1}~(r_{g}~(\frac{1}{r_{3}}-\frac{1}{r_{1}}))^{1/2}
\end{equation}

\section{Conclusion}                                                          
The geodesics of particles in the combined Schwarzschild black hole and a vacuum energy density, are given by ratios of quadrupoly periodic theta functions of two variables. The four moduli of the theta functions are given by ratios of periods of the hyperelliptic integrals of the first kind. These these primitive normal periods may be obtained by integrating between the branch places on the Riemann surface of genus two.\\

Acknowledgement: I would like to thank the University of British Columbia for the use of the computing facilities to produce this document.

\section{References}
[1] Weinberg, S. The Cosmological Constant Problem, Reviews of Modern Physics, Vol. 61 No.1 Jan. 1989.\\
\
[2] Carroll, B.W., Ostlie, D.A., An Introduction to Modern Astrophysics, Addison Westly, 1996.\\
\
[3] Filippenko, A.V., Einstein's Biggest Blunder? High-Redshift Supernovae and the Accelerating Universe, Publications of the Astronomical Society of the Pacific, Vol 113, pp 1441-1448, Dec. 2001\\
\
[4] Efstathlou, G., Evidence for a non-zero /Lambda and a low matter density from a combined analysis of the 2dF Galaxy Redshift Survey and cosmic microwave background anisotropies, Mon. Not. R. Astron. Soc. Vol 330 L29-L35 (2002)\\
\
[5] Neupane, G., Planetary Perturbation with a Cosmological Constant, gr-qc/9902039 (12 Feb 1999)\\
\
[6] Cardona, S.B., Can Interplanetary Measures Bound the Cosmological Constant?  The Astrophysical Journal, Vol. 493, pp. 52-53, (1998 Jan 2001)\\
\
[7] Boughn, S.P., Crittenden, R.G., Cross Correlation of the Cosmic Microwave Background with Radio Sources: Constrains on an Accelerating Universe, Physical Review Letters, Vol 88 No 2, 021302-1\\
\
[8] Whitehouse, S.B., A Possible Explanation of Galactic Velocity Rotation Curves in Terms of a Cosmological Constant, astro-ph/9911485 26 Nov. 1999\\
\
[9] Baker, H.F., Abelian Functions, Abel's theorem and allied theory of theta functions. Cambridge Mathematical Library. First Publication 1897 reissued in 1995\\
\
[10] Baker, H.F. Multiply Periodic Functions, Cambridge University Press, 1907\\
\
[11] Witten, E., Anti-de Sitter Space, Thermal Phase Transition, And confinement in gauge theories. hepth/9803131, IASSNS-HEP-98/21, p.7 (16 Mar 1998)\\
\
[12] Gibbons, G.W.,Hawking,S.W., Cosmological event horizons, thermodynamics, and particle creation, Physical Review D, Vol. 5, No. 10 pp. 2738-2756. (15 May 1977)\\
\
[13] Rindler, W., Essential Relativity, Springer-Verlag, 1997\\
\
[14] Bartlett, J.H. Classical and Modern Mechanics, The University of Alabama Press, 1975 p. 194.\\
\
[15] Landau, L.D., Lifshitz, E.M., Mechanics, Oxford, NYNY, Pergamon Press, 1976 $3^{rd}$ edition, QA 805 L283 1976.\\
\
[16] Jacobi, von  K.G.J., Ueber VIERFACH PERIODISCHEN FUNCTIONEN zweier Variabeln, auf die sich die Theorie der Abel'shen Transcendenten stutzt.(Crelle's Journal fur reine und angewandte Mathematik Bd. 13, 1834.)(Herausgegeben von H. Weber aus dem Franzosischen ubersetzt von A. Witting.Leipzig verlag von wilhelm engelmann 1895. Oftwald's Klaffiker der exakten Wiffenfchaften Nr. 64.\\
\ 
[17] Gauss, C.F., see his diary.\\
\
[18] Abel, N.H., Oeuvres completes I,II, by L. Sylow and S. Lie, Grondal and Son, Christiania, new edition 1881.\\
\
[19] Brioshi, F., Sur l'equation du sixieme degre, Acta mathematica, 12 Imprime le 11 octobre 1888.\\
\
[20] Bolza, O. Darstellung der rationalen ganzen Invarianten der Binarform sechsten Grades durch die Nullwerthe der zugehorigen $\vartheta$-Functionen.\\
\
[21] Konigsberger, Von Hern zu Greifswald, Ueber die Transformation des zweiten Grades fur die Abelsen Functionen erster Ordnug. Jan. 1866.\\
\
[22] Rosenhain, von Georg. Abhandlung uber die FUNCTIONEN ZWEIER VARIABLER mit vier Perioden, welche die Inversen sind der ultra-elliptischen integrale erster Klasse(Herausgegeben von H. Weber aus dem Franzosischen ubersetzt von A. Witting.Leipzig verlag von wilhelm engelmann 1895. Oftwald's Klaffiker der exakten Wiffenfchaften Nr. 65.This paper won the mathematics prize in Paris in the 1850's\\
\
[23] Gopel, A.,''Theoriae transcendetium Abelianarum primi ordinis adumbratio levis,' Journal fur die reine und angewandte Mathematik, Vol 35 (1847)\\
\ 
[24] Forsyth, A.R. Introductory to the theory of Functions of Two Variables, Cambridge at the University Press, 1914. QA F69.\\
\
[25] Drociuk, R.J., On the Complete Solution to the Most General Fifth Degree Polynomial, GM/0005026.
\ 
[26] Forsyth, A.R., Memoir on Theta Functions, particularly those of two variables, communicated by Prof. A. Cayley, Phil. Trans. 1881, p. 783.\\
\end{document}